\documentstyle[aaspp4,tighten,flushrt]{article}

\newcommand{\beq}{\begin{equation}}
\newcommand{\eeq}{\end{equation}}
\newcommand{\beqa}{\begin{eqnarray}}
\newcommand{\eeqa}{\end{eqnarray}}

\def\dD{[\delta_{\rm D}]}

\font\BF=cmmib10

\def\k{{\hbox{\BF k}}}
\def\x{{\hbox{\BF x}}}

\def\q{{\hbox{\BF q}}}

\def\la{\mathrel{\mathpalette\fun <}}
\def\ga{\mathrel{\mathpalette\fun >}}
\def\fun#1#2{\lower3.6pt\vbox{\baselineskip0pt\lineskip.9pt
\ialign{$\mathsurround=0pt#1\hfill##\hfil$\crcr#2\crcr\sim\crcr}}}

\begin{document}
\hfill{\small CITA-98-17, FERMILAB-Pub-98/255-A}

\title{Hyperextended Cosmological Perturbation Theory:\\ 
Predicting Non-linear Clustering Amplitudes}

%
%
\author{Rom\'{a}n Scoccimarro$^{1}$ and Joshua A. Frieman$^{2,3}$}

\vskip 1pc

\affil{${}^1$CITA, McLennan Physical Labs, 60 St George Street,
Toronto ON M5S 3H8, Canada}

\affil{${}^2$NASA/Fermilab Astrophysics Center, Fermi National 
Accelerator Laboratory, \\ P.O. Box 500, Batavia, IL  60510}

\affil{${}^3$Department of Astronomy and Astrophysics, 
University of Chicago, Chicago, IL 60637}

%
\begin{abstract}
%
We consider the long-standing problem of predicting the hierarchical
clustering amplitudes $S_p$ in the {\em strongly } non-linear regime
of gravitational evolution. N-body results for the non-linear
evolution of the bispectrum (the Fourier transform of the three-point
density correlation function) suggest a physically motivated ansatz
that yields the strongly non-linear behavior of the skewness, $S_3$,
starting from leading-order perturbation theory. When generalized to
higher-order ($p>3$) polyspectra or correlation functions, this ansatz
leads to a good description of non-linear amplitudes in the strongly
non-linear regime for both scale-free and cold dark matter models.
Furthermore, these results allow us to provide a general fitting
formula for the non-linear evolution of the bispectrum that
interpolates between the weakly and strongly non-linear regimes,
analogous to previous expressions for the power spectrum.

\end{abstract}
\clearpage 

%
%
\section{Introduction}
\label{intro}
%
%

Analytic understanding of gravitational clustering in the {\it
strongly} non-linear regime remains an elusive goal in the study of
large-scale structure.  While the growth of structure in the weakly
non-linear regime, where the density field contrast $\delta \la 1$, is
well described by cosmological perturbation theory (\cite{JBC93};
\cite{B94}, 1994b; \cite{BGE95}; \cite{GB95}; \cite{JWACB95};
\cite{LJWB95}), corresponding progress has not been made in
understanding clustering when $\delta \gg 1$.  This lack of progress
traces in part to the relative complexity and intractability of the
coupled BBGKY hierarchy of equations for the phase space distribution
functions (Peebles 1980); moreover, with the development of
high-resolution N-body simulations, the non-linear gravitational
evolution can now be tracked directly.  However, simulations provide
only limited physical insight into the complex non-linear phenomena
involved in gravitational clustering.  To the extent possible, one
would like a qualitative and ideally quantitative understanding of
non-linear clustering from first principles.  Analytic models can also
provide a useful check on simulation results, which have their own
intrinsic limitations and uncertainties.

Recently, cosmological perturbation theory has been extended further
into the non-linear regime through the inclusion of next-to-leading
order corrections: one-loop perturbative calculations of the power
spectrum (\cite{MSS92}; \cite{LJBH96}; \cite{SF96}), bispectrum
(\cite{Sco97}; \cite{SCFFHM98}, hereafter SCFFHM) and skewness
(\cite{SF95}; \cite{Sco97}; \cite{FoGa98}) have been found to be in
excellent agreement with N-body results even in the regime where the
variance $\sigma^2 \equiv \langle \delta^2(R)\rangle \sim 1$ for the
bispectrum and as large as $\sigma^2 \sim 10$ for the power spectrum
(where $\delta(R)$ is the density field contrast smoothed through a
window of radius $R$).  However, it is clear that perturbation theory
cannot be pushed successfully beyond this point to calculate
clustering amplitudes in the strongly non-linear regime.  The
perturbation series is at best asymptotic, and it is expected to break
down on small scales, where $\sigma^2 \gg 1$.  More importantly, the
single-stream fluid approximation, upon which the perturbative
solutions are based, must fail on small scales where shell-crossing
occurs, and be replaced with the BBGKY equations.

Nevertheless, non-linear gravitational clustering does appear to
display striking scaling properties, and some success has been
achieved in combining perturbation theory with concepts such as stable
clustering to make predictions about strongly non-linear behavior. The
stable clustering hypothesis states that on very small scales, where
clustering has reached virial equilibrium, the mean relative velocity
between pairs should exactly cancel the Hubble expansion. When coupled
with the equations of motion, this leads to general predictions for
the growth of the hierarchy of $p$-point density correlation functions
(Peebles 1980).  For scale-free initial conditions, i.e., power
spectrum $P_i(k) \sim k^n$, in an Einstein-de Sitter universe (with
$\Omega_m=1$), it follows from self-similarity that the slope of the
{\it non-linear} two-point function, $\xi(r) \sim r^{-\gamma}$, is
related to the spectral index $n$ of the initial perturbations by
$\gamma = 3(n+3)/ (n+5)$ (Davis \& Peebles 1977).  For a range of
initial spectra $n$, the results of N-body simulations are consistent
with this self-similar solution in the strongly non-linear regime,
$\xi \ga 10 - 100$ (Efstathiou et al.  1988; \cite{CBH96}, hereafter
CBH; Jain 1997; \cite{CoPe98}). In current models of structure
formation, such as cold dark matter (CDM) and its variants, the
initial power spectrum is not precisely scale-free. Nevertheless, the
spectral index $n= d\ln P/d\ln k$ generally varies slowly enough with
scale that one expects scaling arguments to provide useful insight
into non-linear clustering.  In this paper, we use such considerations
to study the non-linear evolution of higher-order correlations.

A classic problem in this context is the prediction of the strongly
non-linear hierarchical amplitudes $S_p(R) = \langle
\delta^p(R)\rangle/\langle \delta^2(R) \rangle^{p-1}$.  For scale-free
initial conditions in an Einstein-de Sitter universe, the stable
clustering hypothesis extended to higher-order correlations (Peebles
1980, Jain 1997) implies that the $S_p$ should be constant,
independent of $R$, in the strongly non-linear regime, $\sigma^2 \gg
1$, but it does not say what the amplitudes are.

This result motivated several papers which attempted to calculate the
hierarchical amplitudes from the equations of motion. Observations of
galaxy clustering at small scales and the self-similarity solution of
the BBGKY equations motivate the so-called {\em hierarchical model}
for the connected $p$-point correlation function (Fry 1984b),

\beq
\label{HM}
\kappa_p(\x_1,...,\x_p) \equiv \langle \delta(\x_1),...,\delta(\x_p)\rangle_c
= \sum_{a=1}^{t_p} Q_{p,a} \sum_{\rm labelings}\
 \prod_{\rm edges}^{p-1} \xi_{AB} \equiv Q_p \sum_{a=1}^{t_p} \
\sum_{\rm labelings}\ \prod_{\rm edges}^{p-1} \xi_{AB}. 
\eeq 
The product is over $p-1$ edges that link $p$ galaxies (vertices) $A,
B, ...$, with a two-point correlation function $\xi_{XY}$ assigned to
each edge. These configurations can be associated with `tree' graphs,
called $p$-trees. Topologically distinct $p$-trees, denoted by $a$, in
general have different amplitudes, denoted by $Q_{p,a}$, but those
configurations which differ only by permutations of the labels
1,...,$p$ (and therefore correspond to the same topology) have the
same amplitude.  There are $t_p$ distinct $p$-trees ($t_3=1$, $t_4$=2,
etc., see Fry (1984b) and Bosch\'an, Szapudi \& Szalay 1994) and a
total of $p^{p-2}$ labeled trees.  The {\it hierarchical model}
represents the connected $p$-point functions as sums of products of
$(p-1)$ two-point functions, introducing at each level only as many
extra parameters $Q_{p,a}$ as there are distinct topologies.  In what
we shall call the {\em degenerate hierarchical model}, the amplitudes
$Q_{p,a}$ are furthermore independent of scale and configuration.  In
this case, $Q_{p,a}=Q_p$, and the hierarchical amplitudes $S_p \simeq
p^{p-2}\ Q_p$.  In the general case, the amplitudes $Q_p$ depend on
overall scale and configuration. For example, for Gaussian initial
conditions, in the {\it weakly} non-linear regime, $\sigma^2 \ll 1$,
perturbation theory predicts a clustering pattern that is hierarchical
but not degenerate. 

Using the BBGKY hierarchy and assuming a hierarchical form similar to
Eq.~(\ref{HM}) for the {\em phase-space $p$-point distribution
function}, in the stable clustering limit Fry (1982, 1984) obtained
($p\geq 3$)

\beq
\label{Qfry}
Q_p = Q_{p,a} = \frac{1}{2}\ \Big(\frac{p}{p-1}\Big)\
\Big(\frac{4Q_3}{p}\Big)^{p-2}
; 
\eeq
in this case, different tree diagrams all have the same amplitude, 
i.e., the clustering pattern is degenerate. On the
other hand, Hamilton (1988), correcting an unjustified symmetry
assumption in Fry (1982, 1984),  instead found

\beq
\label{Qham}
Q_{p,{\rm snake}} = Q_3^{p-2}, \ \ \ \ \ Q_{p,{\rm star}}=0 
\eeq 
where ``star'' graphs correspond to those tree graphs in which one
vertex is connected to the other $(p-1)$ vertices, the rest being
``snake'' graphs. Summed over the snake graphs, (\ref{Qham}) yields

\beq
Q_p = \frac{p!}{2}\ \Big(\frac{Q_3}{p}\Big)^{p-2}.
\eeq
Unfortunately, as emphasized by Hamilton (1988), these results are
{\em not} physically meaningful solutions to the BBGKY hierarchy, but
rather a direct consequence of the assumed factorization in {\em
phase-space}.  As a result, this approach leads to unphysical
predictions such as that cluster-cluster correlations are equal to
galaxy-galaxy correlations to all orders.  It remains to be seen
whether physically relevant solutions to the BBGKY hierarchy which
satisfy Eq.~(\ref{HM}) really do exist.  Despite these shortcomings,
the results in Eq.~(\ref{Qfry}) and Eq.~(\ref{Qham}) are often quoted
in the literature as physically relevant solutions to the BBGKY
hierarchy!

The complexity of the BBGKY equations has prompted a phenomenological
approach to the description of correlation functions based on the
hierarchical model. An example is the {\em factorizable hierarchical
model} (Bernardeau \& Schaeffer 1992), which is completely specified
by the star amplitudes. In this case the snake amplitudes at any given
order are determined by the lower order star amplitudes,
$Q_{p,a}=\Pi_i \nu_i^{d_i(a)}$, where $\nu_i$ are the vertex weights
for $i$ lines and $d_i(a)$ is the number of such vertices in diagrams
with topology denoted by $a$. This is analogous to the pattern that
emerges from PT at large scales, although the parameters $Q_{p,a}$ are
in this case taken to be constant, independent of scale and
configuration, as in the spherical model.

A more general framework than the hierarchical model is given by the
{\em scale-invariant} model (Balian \& Schaeffer 1989), in which  the
connected $p$-point function obeys the scaling law 
 
\beq
\label{SIM}
\kappa_p(\lambda \x_1,...,\lambda \x_p) = \lambda^{-(p-1)\gamma}
\ \kappa_p(\x_1,...,\x_p), 
\eeq 
where $\gamma$ is the index of the two-point function, $\xi(r,t) \sim
(r/t^{\alpha})^{-\gamma}$, with $\alpha=4/[3(n+3)]$.  Eq.~(\ref{SIM})
is the self-similar solution to the BBGKY hierarchy, which holds in
the case of stable clustering for scale-free initial conditions in an
Einstein-de Sitter universe (Peebles 1980).  The hierarchical model of
Eq.~(\ref{HM}) satisfies Eq.~(\ref{SIM}), but the latter is more
general and can be satisfied by other functional forms than
Eq.~(\ref{HM}).

The best observational constraints on higher-order galaxy correlation
functions in the non-linear regime currently come from angular
surveys, although this situation will soon change dramatically with
the advent of large redshift surveys such as the Two Degree Field
(2dF) and Sloan Digital Sky Survey (SDSS).  Measurement of the angular
three-point function (Groth \& Peebles 1977) and four-point function
(Fry \& Peebles 1978) in the Lick survey provided the first
observational evidence for the scale-invariant model of
Eq.~(\ref{SIM}).  Moreover, these observations are consistent with the
hierarchical model, with $Q_3 \approx 1.3$ and $Q_4\approx 3$.  More
recently, Gazta\~naga (1994) measured the $S_p$ parameters ($3 \leq p
\leq 9$) in the APM galaxy survey and found good agreement with the
scale-invariant model at small scales.  Szapudi \& Szalay (1998)
investigated the third- and fourth-order cumulant correlators in the
APM and found $Q_3 \approx 1$ and $Q_4\approx 3$, in good agreement
with the Lick survey results.  On the other hand, when decomposing the
average four-point amplitude $Q_4$ into the two different topologies,
Szapudi \& Szalay (1998) (see also Szapudi et al. 1995) found
$R_a=Q_{4,{\rm snake}}=3.7$ (for the snake topology) and $R_b=Q_{4,{\rm
star}}=0.8$ (star topology), whereas Fry \& Peebles (1978) obtained
$R_a=2.5 \pm 0.6$ and $R_b=4.3 \pm 1.2$ for the Lick survey.  One must
keep in mind, however, that these two measurements differ in the type
of four-point configurations they considered (that is, they did not
test the hierarchical ansatz in its full generality), so it remains
possible that a finer measurement of the four-point function as a
function of configuration could reconcile these discrepancies.
Furthermore, in the EDSGC survey (which is based on a subset of the
same photographic plates as used in the APM), Szapudi, Meiksin \&
Nichol (1996) found that $Q_3=2.0$, $Q_4=7.3$, substantially higher
than in the APM (see Szapudi \& Gazta\~naga 1998 for a detailed
comparison of higher order correlations in these two surveys).  The
upcoming 2dF and SDSS redshift surveys will be able to probe
higher-order correlations with unprecedented accuracy and should help
resolve these issues (\cite{CSS98}).  We note, however, that testing
the hierarchical model in these cases will require redshift
distortions to be taken into account; in particular, the large
internal velocity dispersion of galaxy clusters leads to a strong
configuration-dependence of the $Q_p$ amplitudes at small scales
(\cite{SCF98}).  In fact, this expected violation of the (degenerate)
hierarchical ansatz is seen in the three-point function measured in
the Las Campanas Redshift Survey (Jing \& B\"orner 1998).

In this paper, we combine cosmological perturbation theory with the
observed scaling behavior of the three-point function in N-body
simulations to extract predictions for the non-linear clustering
amplitudes $S_p$ for scale-free and CDM initial power spectra.  The
end result is a simple analytic expression for $S_p$ which appears to
be valid at $\sigma^{2} \ga 10$ when compared with N-body results.
Along the way, we also provide fitting formulae for the non-linear
evolution of the bispectrum, the Fourier transform of the three-point
function, in the spirit of the two-point results of Hamilton et al.
(1991), Jain, Mo \& White (1995), and Peacock \& Dodds (1994, 1996).

In recent work, Colombi et al. (1997) take a very different approach
to `extending' perturbative expressions for the $S_p$ parameters into
the non-linear regime. In leading order perturbation theory (hereafter
PT), valid for $\sigma^{2} \ll 1$, the $S_p$ are functions of the
linear spectral index, $n = -3 - d\ln \sigma^2_\ell(R) / d\ln R$,
where $\sigma^2_\ell(R)$ denotes the linear variance at smoothing
scale $R$ (Bernardeau 1994).  Colombi, et al. find that the non-linear
evolution of the $S_p$ can be fit with the {\it same} PT expressions
provided they let the spectral index $n$ appearing therein become a
{\em free parameter} as a function of the non-linear variance, $n
\rightarrow n_{\rm eff}(\sigma^2)$.  (Note that $n_{\rm eff}$ is {\it
not} the slope of the non-linear power spectrum.)  The function
$n_{\rm eff}(\sigma^{2})$, which depends on the initial spectrum, is
extracted from, e.g., the measurement of $S_3(\sigma^2)$ in an N-body
simulation, assuming that the perturbative expression for the skewness
remains valid throughout the non-linear regime, i.e., by fitting the
N-body results to $S_3(\sigma^2) = (34/7)-(n_{\rm eff}+3)$.
Remarkably, they find that the extracted $n_{\rm eff}(\sigma^{2})$,
when substituted into the PT expressions for the $S_{p}$ for $p>3$,
provide a good fit to the non-linear (N-body) evolution of the
higher-order moments as well. Similar behavior was noted for the EDSGC
moments by Szapudi, Meiksin, and Nichol (1996).  This {\em extended
perturbation theory} (EPT) in principle provides a complete
phenomenological description of one-point density field statistics,
from linear to non-linear scales.  (We describe it as a
phenomenological model, because it is not a systematic development of
perturbation theory for the non-linear regime; rather it is based upon
the observation that the pattern of the $S_p$ in the non-linear regime
is related by a single parameter to the pattern in the weakly
non-linear (perturbative) regime.)  Note that EPT does not give a
predictive prescription for the non-linear $S_p$ for arbitrary initial
conditions: if the initial conditions are changed, a new N-body
simulation must be run in order to fit the function $n(\sigma^2)$. In
addition, it does not explictly describe the evolution of the
multi-point functions.

The model discussed below, which we have dubbed ``hyperextended
perturbation theory" (HEPT), also contains phenomenological elements:
it does not provide a rigorous, first-principles calculation of the
strongly non-linear clustering amplitudes.  However, it is based upon
a simple physical picture of the non-linear evolution of the $N-$point
functions that appears to hold quite generally. Thus, unlike extended
perturbation theory, hyperextended perturbation theory yields an
analytic prediction for the non-linear $S_p$ for arbitrary initial
conditions.

This paper is organized as follows.  Section~2 reviews the non-linear
evolution of the bispectrum in PT and in N-body simulations and
physically motivates the main ideas behind HEPT. In Section~3 we
describe HEPT and present analytic results for the $S_p$ parameters in
the non-linear regime, comparing them to measurements in numerical
simulations.  A fitting formula for the non-linear evolution of the
bispectrum from the weakly to the strongly non-linear regime is given
in Section~4.  Finally, Section~5 presents our conclusions.

%
\section{Extracting the Essence of HEPT: Non-Linear Evolution of the Bispectrum}
\label{bispec}
%

We are interested in predicting the non-linear behavior of the
higher-order correlations.  For this purpose, we review here the
non-linear evolution of the bispectrum, the Fourier transform of the
three-point spatial correlation function, which is the lowest-order
correlation function sensitive to phase information.

It is useful to first recall the nomenclature of higher order correlations 
in Fourier space. Defining the Fourier transform of the density 
contrast field by

\beq
\delta(\k) = \int \frac{d^3x}{(2\pi)^3}\ e^{-i\k \cdot \x}\ \delta(\x)~,
\eeq
the p$^{\rm th}$ order polyspectrum, $B_p$, is defined by
the expectation value

\beq
\langle \delta (\k_1) \dots \delta (\k_p) \rangle = \dD_p \ 
B_{p}(\k_{1},\ldots, \k_{p}).
\eeq
where $\dD_p \equiv \delta_D(\k_1+\dots +\k_p)$. For $p=2$, this
defines the power spectrum, and for $p=3$ the bispectrum. It is also
convenient to define the $Q_{p}$ hierarchical amplitudes in Fourier
space [see Eq.~(\ref{HM})] by ($p>2$)

\beq
\label{Qp}
Q_{p} \equiv \frac{B_{p}(\k_{1},\ldots, \k_{p})}{ \sum_{a=1}^{t_p} \
\sum_{\rm labelings} \prod_{\rm edges}^{p-1} P(k_{\rm AB}), } 
\eeq
where $P(k)$ is the power spectrum, and the sum in the denominator is
over all the tree diagrams, as in Eq.~(\ref{HM}). For example, $Q_{3}=
B_{3}/(P_{1}P_{2}+P_{2}P_{3}+P_{3}P_{1})$, $Q_{4}=
B_{4}/[P_{1}P_{2}P_{3} + ({\rm 3~permutations}) + P_{1}P_{12}P_{4}+
({\rm 11~permutations})]$, and so on. Here $P_{i}\equiv P(\k_i)$ and
$P_{ij}\equiv P(\k_i+\k_j)$. The hierarchical $S_p(R)$ parameters at
smoothing scale $R$ are the one-point counterpart of the $Q_p$
amplitudes smoothed over a window of radius $R$:

\beq
\label{Sp}
S_p(R) \equiv \frac{\langle \delta^p(R)\rangle_c}{\langle
\delta^2(R)\rangle^{p-1}}  =
\frac{\int B_{p}(\k_{1},\ldots, \k_{p})\ W_1 \ldots W_p\ \dD_p \ d^3k_1
\ldots d^3k_p}{[ \int P(k) W(kR)^2 d^3k]^{p-1}},  
\eeq
where $W_i \equiv W(k_iR)$, with $W(kR)$ the top-hat window function
in Fourier space.

The non-linear evolution of $Q_{3}$ displays the main features
expected to hold for higher-order polyspectra as well.  As its
behavior has been rather thoroughly studied in both PT and N-body
simulations, we shall use the three-point function as a model to
extrapolate to higher-order polyspectra.  The $p>3$-order correlation
functions become increasingly difficult to measure in numerical
simulations for increasing $p$, so less is known about their
non-linear evolution (however, see the studies of the four-point
function by, e.g., \cite{Bromley94}, \cite{SuMa94}, \cite{MM98}). On
the other hand, there is partial information from studies of the
evolution of one-point statistics such as the $S_{p}$ up to $p=10$
(\cite{BGE95}; CBH).

There are basically three qualitatively distinct regimes for the
non-linear evolution of $Q_{3}$:

a) {\em Tree-Level}.  At large scales, leading-order (tree-level) PT
provides an excellent description of gravitational clustering.  In
this regime, the $Q_{p}$ amplitudes in Eq.~(\ref{Qp}) are independent
of the overall amplitude of the power spectrum. For scale-free initial
conditions, $P_k(k) \sim k^n$, the PT $Q_p$ are scale-invariant but
configuration-dependent, that is, they depend on the ratios
$k_{i}/k_{j}$ and the angles $\k_{i}\cdot \k_{j}/( k_{i}k_{j})$ (Fry
1984b).  The resulting $S_{p}$ parameters depend on scale only through
the spectral index at the smoothing scale and its derivatives
(\cite{B94}), therefore they are constants for scale-free initial
conditions.

b) {\em One-Loop}. By carrying the perturbative expansion for the
$p-$point functions to next-to-leading order in the density contrast
field, one includes the one-loop corrections to the tree-level
results.  One-Loop PT describes the dependence of $Q_{p}$ and $S_{p}$
on the amplitude of the power spectrum that appears at intermediate
scales.  Depending on the initial spectral index, one-loop corrections
tend to enhance ($n_{\rm eff}~>~-1.4$) or reduce ($n_{\rm
eff}~<~-1.4$) the tree-level configuration dependence of $Q_3$
(Scoccimarro 1997).  For equilateral triangle configurations, one-loop
PT describes the rise that connects the tree-level perturbative
amplitude, $Q_{3}=4/7$, to the non-linear saturation value at small
scales (SCFFHM).  Similarly, one-loop corrections to the skewness
$S_{3}$ describe the transition from its tree-level value to the
non-linear regime (Scoccimarro 1997). Extension of this result within
the spherical collapse approximation shows that one-loop corrections
describe this transition region for $S_{p}$ with $p>3$ as well
(\cite{FoGa98}).

c) {\em Saturation}.  At small scales, in the strongly non-linear
regime, numerical simulations show that the $S_{p}(R)$ parameters
reach a plateau, nearly independent of scale $R$; we refer to this
behavior as ``saturation''.  The hierarchical amplitude $Q_{3}$ in
this regime becomes constant to a good approximation, nearly
independent of configuration (SCFFHM).  Note, however, that it is
still not settled whether the $S_p$ show small deviations from
scale-invariant behavior in the strongly non-linear regime (CBH;
\cite{MBMS97}).  From a theoretical point of view, as we discussed
above, the only prediction in the strongly non-linear regime is from
stable clustering, which implies that the $S_p$ should be constant for
scale-free initial conditions in an Einstein-de Sitter model.  As for
the $Q_{p}$ parameters, however, stable clustering only constrains
them to be scale-invariant (not necessarily hierarchical); in
particular, for the bispectrum this implies

\beq
\label{Sinv}
B(k_{1},k_{2},k_{3})= P(k_{1}) P(k_{2}) \ {\cal S}(r_{12},\theta_{12})
+ P(k_{2}) P(k_{3})\ {\cal S}(r_{23},\theta_{23}) + P(k_{3}) P(k_{1})\
{\cal S}(r_{31},\theta_{31}) . 
\eeq 
Here, ${\cal S}(r,\theta)$ is some arbitrary function (symmetrized
over $k_1$ and $k_2$) of the ratios $r_{ij}\equiv k_{i}/k_{j}$ and
angles $\cos \theta_{ij} \equiv (\k_i \cdot \k_j)/(k_i k_j)$; in the
hierarchical model, ${\cal S}(r,\theta)=Q_{3}=$ constant. In order to
preserve self-similarity, the time dependence of the bispectrum is
driven by that of the power spectrum.  The form in Eq.~(\ref{Sinv})
generalizes to higher-order polyspectra, as in Eq.~(\ref{HM}), where
now each amplitude $Q_{p,a}$ becomes a different function of the
ratios $r_{ij} $ and angles $\theta_{ij}$.  This is exactly the
behavior of higher-order correlations in tree-level PT (Fry 1984b).

Since symmetries do not require a hierarchical three-point function in
the non-linear regime, the N-body results suggest that the physics of
gravitational clustering leads to ${\cal S}(r,\theta)$ being nearly
constant, i.e., the saturation value for $Q_{3}$ is only weakly
dependent on $r$ and $\theta$.  To see how this might arise, consider
the interpretation of the Fourier hierarchical amplitude $Q_3$ in
tree-level PT. In this case, the function ${\cal S}(r,\theta)$ is
given by

\beq
\label{SPT}
{\cal S}(r,\theta)= \frac{10}{7} + \cos \theta\
\Big(r+\frac{1}{r} \Big) + \frac{4}{7} \cos^2 \theta,
\eeq

\noindent obtained from second-order PT (Fry 1984b). The configuration
dependence through $r$ and $\theta$ in Eq.~(\ref{SPT}) comes from
gradients of the density and velocity fields in the direction of the
flow: the dependence on configuration arises from the anisotropy of
structures and flows generated by the physics of gravitational
instability (Scoccimarro 1997). Eq.~(\ref{SPT}) implies that the
hierarchical amplitude $Q_3$ is maximum for collinear configurations
($\cos \theta =\pm 1$) and minimum for isosceles configurations (where
two sides of the triangle are equal). This reflects the fact that
gravitational instability generates large-scale flows mostly parallel
to density gradients, which enhances collinear configurations in
Fourier space. On the other hand, on small scales, where virialization
leads to substantial velocity dispersion, this picture suggests that
the function ${\cal S}(r,\theta)$ should approach a constant. That is,
non-collinear configurations become more probable than at large
scales: the loss of coherence between structures and flows implies
that there is no reason to expect some configurations to be enhanced
over others.

Figure 1 illustrates these points for a CDM simulation, done by
Couchman, Thomas \& Pearce (1995) with an adaptive P$^{3}$M code that
involves $128^{3}$ particles in a box of length $ 100 \, h^{-1} \,
{\rm Mpc} $ ($ h \equiv H_0/100 \, {\rm km \, s^{-1} \, Mpc^{-1}} $,
where $H_0$ is the Hubble constant).  These simulation data are
publicly available through the Hydra Consortium Web page ({\sf
http://coho.astro.uwo.ca/pub/consort.html}) and correspond to an
$\Omega_m=1$ model, with linear CDM power spectrum characterized by a
shape parameter $\Gamma= 0.25$, and normalization $\sigma_8=0.64$
($\tau$CDM).  The top right panel shows the dependence on scale of the
hierarchical amplitude $Q_{\rm EQ}$ for equilateral triangles in
Fourier space.  At large scales (small $k$), $Q_{\rm EQ}$ approaches
the tree-level PT value shown in the dashed line, $Q_{\rm
EQ}=4/7=0.57$.  At intermediate scales, $Q_{\rm EQ}$ rises ({\em
one-loop regime}), eventually flattening at small scales ({\em
saturation}). As shown in the figure, one-loop PT provides a good
description of the transition regime but clearly breaks down in the
saturation regime.  The lower panels in Fig. 1 illustrate the
dependence of the saturation value of $Q_3$ on configuration, for
$k_1/k_2=2$ (bottom right) and $k_1/k_2=3,4$ (bottom left).  Although
self-similarity considerations do not strictly apply to
scale-dependent spectra such as the $\tau$CDM model, there is
remarkably little dependence on configuration in the strongly
non-linear regime (perhaps a slight decrease of $Q_3$ with increasing
$k_1/k_2$ ratio), which confirms the expectations based on the
physical picture discussed above.

An important observation from Fig.~1 is that the saturation value of
$Q_3$ is in good agreement with the collinear configuration value
$Q^{\rm TL}(0,\pi)$ given by tree-level PT (maxima of the dashed
curves, at $\theta=0,\pi$).  Although the dashed curves correspond to
$k_1/k_2=2$ configurations, the tree-level collinear amplitude $Q^{\rm
TL}(0,\pi)$ is very insensitive to the ratio $k_1/k_2$.  In fact, for
spectral indices $n=-2,0$, $Q^{\rm TL}(0,\pi)$ is independent of the
ratio $k_1/k_2$.  For these spectra, and in general to an excellent
approximation, collinear configurations are invariant under arbitrary
scaling of the different triangle sides.  This property singles out
these configurations, and we will take advantage of it to predict
higher-order correlation amplitudes $S_p$ in the non-linear regime.

%
\section{Hyperextended Perturbation Theory}
\label{hept}
%

We shall now assume that clustering does reach approximate
scale-invariance (saturation) at small scales (at least locally for
CDM spectra).  Based on the results above, we shall propose a
physically motivated ansatz that allows one to calculate the $S_p$
parameters in the non-linear regime purely from knowledge of
tree-level PT.

The discussion above shows that collinear configurations play a
special role in gravitational clustering.  They correspond to matter
flowing parallel to density gradients, thus enhancing clustering at
small scales until eventually giving rise to bound objects that
support themselves by velocity dispersion (virialization).  {\em We
thus conjecture that the ``effective'' $Q_p$ clustering amplitudes in
the strongly non-linear regime are the same as the weakly non-linear
(tree-level PT) collinear amplitudes}, as shown in Fig.~1 to hold for
three-point correlations.  We name this ansatz ``hyperextended
perturbation theory'' (HEPT), since it borrows PT ideas valid at the
largest scales to predict the behavior of clustering at small scales.

Note that by effective amplitudes $Q_p^{\rm eff}$ we mean the overall
magnitude of $Q_p$: it is possible that $Q_p$, for $p>3$, although
scale-invariant, is a function of configuration (as, e.g., in a
non-degenerate hierarchical model, in which different topologies have
different amplitudes $Q_{p,a}$).  To calculate the resulting $S_p$
parameters, we assume that $S_p \simeq p^{p-2}\ Q_p^{\rm eff}$, that
is, the $S_p$ are given by the typical configuration amplitude
$Q_p^{\rm eff}$ times the total number of labeled trees, $p^{p-2}$.
In practice, there is a small correction to this formula due to
smoothing, which we neglect (\cite{BSS94}).

To obtain quantitative predictions, we must specify which tree-level
collinear configurations we use to calculate. As $p$ increases from 3,
there is a growing number of collinear configurations, with different
ratios $r_{ij}=k_i/k_j$. However, as noted above, in tree-level PT
different collinear amplitudes depend only very weakly on $r_{ij}$, as
in the $p=3$ case. In what follows, we choose the $\k_1= \ldots =
\k_{p-1} =\q$, $\k_p=-(p-1) \q$ configuration to calculate $Q_p$.  The
resulting non-linear $S_p$ amplitudes follow from tree-level PT

\beq
\label{s3}
S_3^{\rm sat}(n) = 3\ Q_3^{\rm sat}(n) = 3\ \frac{4-2^n}{1+2^{n+1}},
\eeq

\beq
\label{s4}
S_4^{\rm sat}(n) = 16\ Q_4^{\rm sat}(n) = 8\ \frac{54-27\ 2^n+2\ 3^n +
6^n}{(1+6\ 2^n + 3\ 3^n + 6\ 6^n)}.
\eeq

\beq
\label{s5}
S_5^{\rm sat}(n) = 125\ Q_5^{\rm sat}(n) = 125\ \frac{1536-1152\ 2^n+128\
3^n+66\ 4^n+64\ 6^n-9\ 8^n -2\ 12^n -24^n}{6\ (1+12\ 2^n+12\ 3^n+16\
4^n+24\ 6^n+24\ 8^n+12\ 12^n+24\ 24^n)},
\eeq

\noindent where $n$ is the spectral index, obtained from $(n+3) \equiv
- d\ln \sigma^2_\ell(R) / d\ln R$, where $\sigma^2_\ell(R)$ denotes
the linear variance at smoothing scale $R$.  One can check that these
$Q_p$ amplitudes satisfy the constraint that cluster-cluster
correlations are stronger than galaxy-galaxy correlations,

\beq
\label{CC>GG}
Q_p \geq \frac{1}{2} \Big(\frac{p-1}{p}\Big)^{p-3}\ Q_{p-1} \geq
\ldots \geq \frac{p!}{2^{p-1} p^{p-2}} 
\eeq 
(\cite{HaGo88}), as long as $n \la 0.75$, well within the physically
interesting range. The constraint that the one-point probability
distribution function is positive definite leads to $Q_p \geq p^{p-2}$
(Fry 1984), which is weaker than Eq.~(\ref{CC>GG}) and thus
automatically satisfied.

Figure 2 shows a comparison of these predictions with the numerical
simulation measurements of CBH for scale-free initial conditions in an
Einstein-de Sitter universe. These N-body simulations used a tree code
(Hernquist, Bouchet \& Suto 1991) to evolve $64^3$ particles in a
cubic box with periodic boundary conditions. The plotted values
correspond to the measured value of $S_p$ when the non-linear variance
$\sigma^2=100$ (see CBH, table 3). The error bars denote the
uncertainties due to the finite-volume correction applied to the raw
$S_p$ values. We see that the N-body results are generally in good
agreement with the predictions of HEPT, Eqs.~(\ref{s3}), (\ref{s4})
and (\ref{s5}).  The small discrepancy at $n=-2$ may be due to the
excessive large-scale power in this model: in this case, the
finite-volume corrections to the $S_p$ measured in the simulations are
quite large and thus uncertain.

Figure 3 shows a similar comparison of HEPT with numerical simulations
(\cite{CBS94}) in the non-linear regime for the standard CDM model
(with $\Gamma =0.5$, $\sigma_8=0.34$). These N-body simulations used a
P$^3$M code to evolve $64^3$ particles in a box 64 Mpc on a side. The
simulation error bars were estimated as in Fig.~2. The agreement
between the N-body results and the HEPT predictions is very
encouraging indeed. The change in the HEPT saturation value of the
$S_p$ with scale is due to the scale-dependence of the linear CDM
spectral index, and follows the N-body results from $\sigma^2 \simeq
10$ to $\sigma^2 = 300$, where stable clustering is approximately
expected to hold.

Is interesting to note that for $n=0$, HEPT predicts $S_p=(2p-3)!!$,
which agrees exactly with the excursion set model developed by Sheth
(1998) for white-noise Gaussian initial fluctuations. In this case,
the one-point probability distribution function for the density field
yields an inverse Gaussian distribution, which has been shown to agree
very well in the non-linear regime when compared to numerical
simulations (Sheth 1998). This remarkable agreement between HEPT and
the excursion set model deserves further work to understand the
relation, if any, between these seemingly very different approaches
to the description of statistics in the non-linear regime.

From these comparisons, we conclude that HEPT provides a very good
description of one-point statistics in the strongly non-linear
regime. Note that this ansatz, although physically motivated, has not
been rigorously proved from a theoretical point of view. Such a proof
may be extremely difficult to achieve, due to the complexity of
gravitational instability in the strongly nonlinear regime.  However,
the success of HEPT suggests that there is a deep connection between
the physics of the non-linear regime and large-scale clustering. Such
a connection was recently noted by Colombi et al. (1997) in
formulating EPT, although the reason for it was not identified.

%
\section{A Fitting Formula for the Bispectrum}
\label{fit}
%

In this section we provide a fitting formula that describes the
non-linear evolution of the bispectrum as a function of scale,
analogous to previous results in the literature for the power
spectrum. The expression below interpolates between the perturbative
results at tree-level (Fry 1984b) and one-loop (Scoccimarro 1997;
SCFFHM) and the saturation regime at small scales as studied by
numerical simulations (\cite{FMS93}; SCFFHM) and described by HEPT. In
order to describe both the weakly and strongly non-linear regimes, we
take the following form for the bispectrum, inspired by the PT
expression:

\beq
\label{Bfit}
B(k_1,k_2,k_3)= 2 F_2^{\rm eff}(\k_1,\k_2) \ P(k_1)\  P(k_2) + {\rm
permutations} ,
\eeq
where $P(k)$ is the non-linear power spectrum (obtained, e.g., from 
the fitting formulae of Peacock \& Dodds 1996), and the effective 
kernel 
\beq
\label{F2fit}
F_2^{\rm eff}(\k_1,\k_2)= \frac{5}{7}\ a(n,k_1)a(n,k_2) + \frac{1}{2}
\frac{\k_1 \cdot \k_2}{k_1\ k_2}
\Big(\frac{k_1}{k_2}+\frac{k_2}{k_1}\Big) \ b(n,k_1)b(n,k_2) +
\frac{2}{7} \frac{(\k_1 \cdot \k_2)^2}{k_1^2\ k_2^2}\
c(n,k_1)c(n,k_2). 
\eeq 

\noindent When the functions $a=b=c=1$, we recover the tree-level PT
expression for the bispectrum; on the other hand, for $a^2=(7/10)\
Q_3^{\rm sat}$ and $b=c=0$, we recover the results of HEPT for the
strongly non-linear regime.  The functions $a(n,k)$, $b(n,k)$, and
$c(n,k)$ are chosen to interpolate between these two regimes according
to the one-loop PT and N-body results (Scoccimarro 1997; SCFFHM). This
yields ($-2\leq n \leq 0$)
\beqa
\label{coefit1}
a(n,k) &=& \frac{1+ [0.7\ Q_3^{\rm sat}(n)]^{1/2}\ (kR_0)^{n+6}}{1 +
(kR_0)^{n+6}}, \\ & & \nonumber \\
\label{coefit2}
b(n,k) &=& \frac{1+ 0.2\ (n+3)\ (kR_0)^{n+3}}{1 + (kR_0)^{n+3.5}},
\\ & & \nonumber \\
\label{coefit3}
c(n,k) &=& \frac{1+ 4.5/[1.5+(n+3)^4]\ (kR_0)^{n+3}}{1 + (kR_0)^{n+3.5}},
\eeqa

\noindent where the saturation value for the reduced bispectrum
$Q_3^{\rm sat}(n)$ is given by Eq.~(\ref{s3}), and $R_0$ is the value
of the correlation length in linear theory for Gaussian smoothing,
i.e., $\sigma^2_\ell(R_0) \equiv 1$ with a Gaussian window
function. If desired, we can allow for residual scale-dependence of
the saturation value by taking $Q_3^{\rm sat} \rightarrow Q_3^{\rm
sat}\ (\sigma^2/100)^{f(n)}$ (\cite{EPT}; see also CBH). Here, we will
focus on scale-free models and assume that scaling is achieved on the
smallest scales.

Figure 4 shows the fitting formula for the hierarchical three-point
amplitude $Q_3$ (solid curves) as a function of angle $\theta$ between
$\k_1$ and $\k_2$ for configurations with $k_1/k_2=2$
configurations. These configurations are studied at different scales
characterized by the value of $k_1 R_0$.  The power spectrum amplitude
$\Delta(k) = 4\pi k^3 P(k)$ provides a measure of the degree of
non-linearity on these scales. The numerical simulation measurements
shown here are taken from Figs.~1, 2, and 3 of SCFFHM. These N-body
results correspond to $256^3$ PM simulations run by E.~Hivon; the
bispectrum measurements were done by S.~Colombi, according to the
scheme outlined in Appendix A of SCFFHM. In order to obtain $Q_3$ from
Eq.~(\ref{Bfit}) we have used Eq.~(\ref{Qp}) and used the {\em linear}
power spectrum $P(k) \propto k^n$; due to cancellations between the
numerator and denominator, using the non-linear power spectrum (e.g.,
as given by \cite{PD96}) does not change $Q_3$ appreciably, though it
does affect the bispectrum itself.  Figure~5 shows a plot of the
evolution of the equilateral hierarchical amplitude $Q_{\rm EQ}$ as a
function of scale for $n=-1.5$ scale-free initial conditions. The
results in Figs.~4 and 5 were found to be typical of the accuracy of
the fitting formula, which is of order $15\%$ for the scale-free
models considered in SCFFHM. Generalization of
Eqs.~(\ref{coefit1})-(\ref{coefit3}) to CDM spectra is under way and
will be reported elsewhere.

Note that at intermediate scales, the functions $a(n,k)$, $b(n,k)$,
and $c(n,k)$ in Eq.~(\ref{F2fit}) break the scale invariance of $Q$,
as required in the one-loop regime. We have chosen the overall scale
dependence to be separable, in order to simplify applications of the
fitting formula to large-scale structure calculations. For example,
the skewness may be interpolated by

\beq
\label{S3fit}
S_3(R)=  \frac{30}{7}\ \frac{\sigma_a^4}{\sigma^4} + \frac{4}{7}\
\frac{\sigma_c^4}{\sigma^4} - \frac{\sigma_b^4}{\sigma^4}\ (n+3),
\eeq

\noindent where
\beq
\sigma_j^2(R) \equiv \int d^3k\ P(k)\ j(n,k)\ W^2(k R),
\eeq

\noindent for $j=a,b,c$.  In this way, only straightforward
one-dimensional numerical integrations are required to calculate $S_3$
at any scale $R$, using, e.g., the non-linear power spectrum from
Peacock \& Dodds (1996).  Figure 6 shows an application of this result
for $n=-1$ scale-free initial conditions (solid curve), compared to
numerical simulation measurements by CBH. Different symbols denote
different expansion factors, $a=2.5$ (triangles), $a=6.4$ (squares),
and $a=16$ (pentagons), where initial conditions were set at $a=1$.
Error bars, not shown for clarity, are of the order of $20\%$ (CBH);
thus the results of Eq.~(\ref{S3fit}) are well within the N-body
uncertainties. It seems, however, that the fit in Eq.~(\ref{S3fit})
lies consistently above the N-body results. This is to be expected, at
least up to $\sigma^2 \approx 1$, since the numerical simulation
measurements are affected by transients from the Zel'dovich
approximation used to set up the initial conditions (Scoccimarro
1998). In fact, for the $a=2.5$ output (triangles) in Fig.~6, the
$\sigma^2 \rightarrow 0$ limit including transients corresponds to
$S_3=2.39$ instead of the tree-level PT value $S_3=2.86$ shown by the
dashed line.

%
%
\section{Conclusion}
\label{conclusions}
%
%

We have proposed a simple ansatz, called Hyperextended Perturbation
Theory (HEPT), to calculate clustering amplitudes, such as the
hierarchical $S_p$ parameters, in the {\em strongly} non-linear
regime.  Based on N-body studies of scale-free and CDM models, the
HEPT ansatz appears to be valid for all initial conditions of physical
interest and has the advantage that it contains no free
parameters. The proposal is based on extrapolating tree-level PT to
the non-linear regime by using configurations that correspond to
matter flowing parallel to density gradients.

The similarity between the hierarchy of tree-level and strongly
non-linear $S_p$ parameters was pointed out by Colombi et al. (1997),
who noticed that by using the tree-level expressions for the $S_p$
with an arbitrary spectral index they could fit the whole hierarchy of
$S_p$ parameters in the non-linear regime. However, no explanation for
such a remarkable coincidence was given. Within the HEPT framework,
this arises naturally as a consequence of identifying the most
important configurations that drive gravitational collapse. This
provides a valuable tool for quantitative understanding of the
behavior of higher-order correlations in the non-linear regime, based
on simple physics borrowed from perturbation theory. At the same time,
we caution that these results are only meant to provide insight into
non-linear gravitational clustering: in the real universe, the
clustering of luminous galaxies on these scales will be strongly
affected by non-gravitational phenomena as well, such as gas
dissipation, star formation, shocks, etc.

We have also provided a fitting formula for the evolution of the
bispectrum in scale-free models from the weakly to the strongly
non-linear regime.  This expression interpolates between the
perturbative results on large scales and the saturation behavior
observed in simulations at small scales. This result should be useful
for applications in large-scale structure calculations. For example,
using our results to obtain the skewness $S_3$ as a function of scale
and initial spectrum, one can implement the EPT framework of Colombi
et al. (1997) to predict the rest of the one-point cumulants, $S_p$,
for $p>3$ at any scale.

\acknowledgements We thank Francis Bernardeau, Andrew Hamilton, and
Istv\'an Szapudi for useful discussions and St\'ephane Colombi for
providing the numerical simulation measurements in CBH and Colombi,
Bouchet, \& Schaeffer (1994). We also thank Ravi Sheth for pointing
out the agreement of HEPT with the excursion set model for white-noise
Gaussian fluctuations. This research was supported in part by the DOE
at Chicago and Fermilab and by NASA grant NAG5-7092 at Fermilab.  The
$\tau$CDM simulations analyzed in this work were obtained from the
data bank of cosmological $N$-body simulations provided by the Hydra
consortium ({\sf http://coho.astro.uwo.ca/pub/data.html}) and produced
using the Hydra $N$-body code (Couchman, Thomas, \& Pearce 1995).

\clearpage
%
%


\clearpage



\begin{figure}[t!]
\centering
\centerline{\epsfxsize=18truecm\epsfysize=18truecm\epsfbox{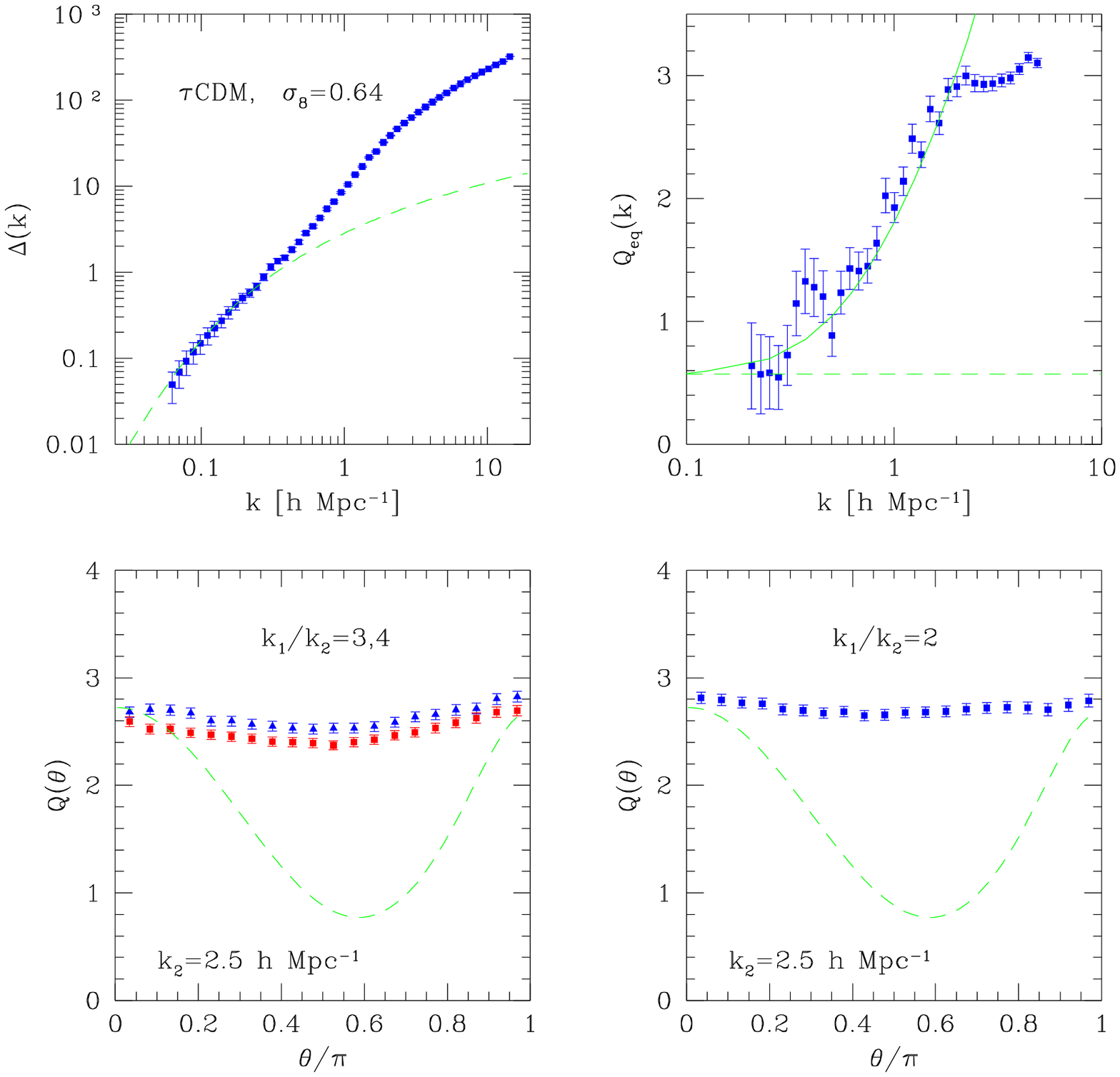}}

\caption{The upper left panel shows the power spectrum,
$\Delta(k)=4\pi k^3 P(k)$, for the $\tau$CDM model
($\sigma_8=0.64$). Other panels show the hierarchical three-point
Fourier amplitude $Q_3$ for equilateral triangles (top right) and for
configurations with different $k_1/k_2$ ratios (bottom) in the
non-linear regime (in the lower left, triangles correspond to
$k_1/k_2=3$, squares to $k_1/k_2=4$). Dashed curves show the
predictions of tree-level PT. The solid curve in upper right panel
shows the prediction of one-loop PT.}
\label{fig1}
\end{figure}

\begin{figure}[t!]
\centering
\centerline{\epsfxsize=18truecm\epsfysize=18truecm\epsfbox{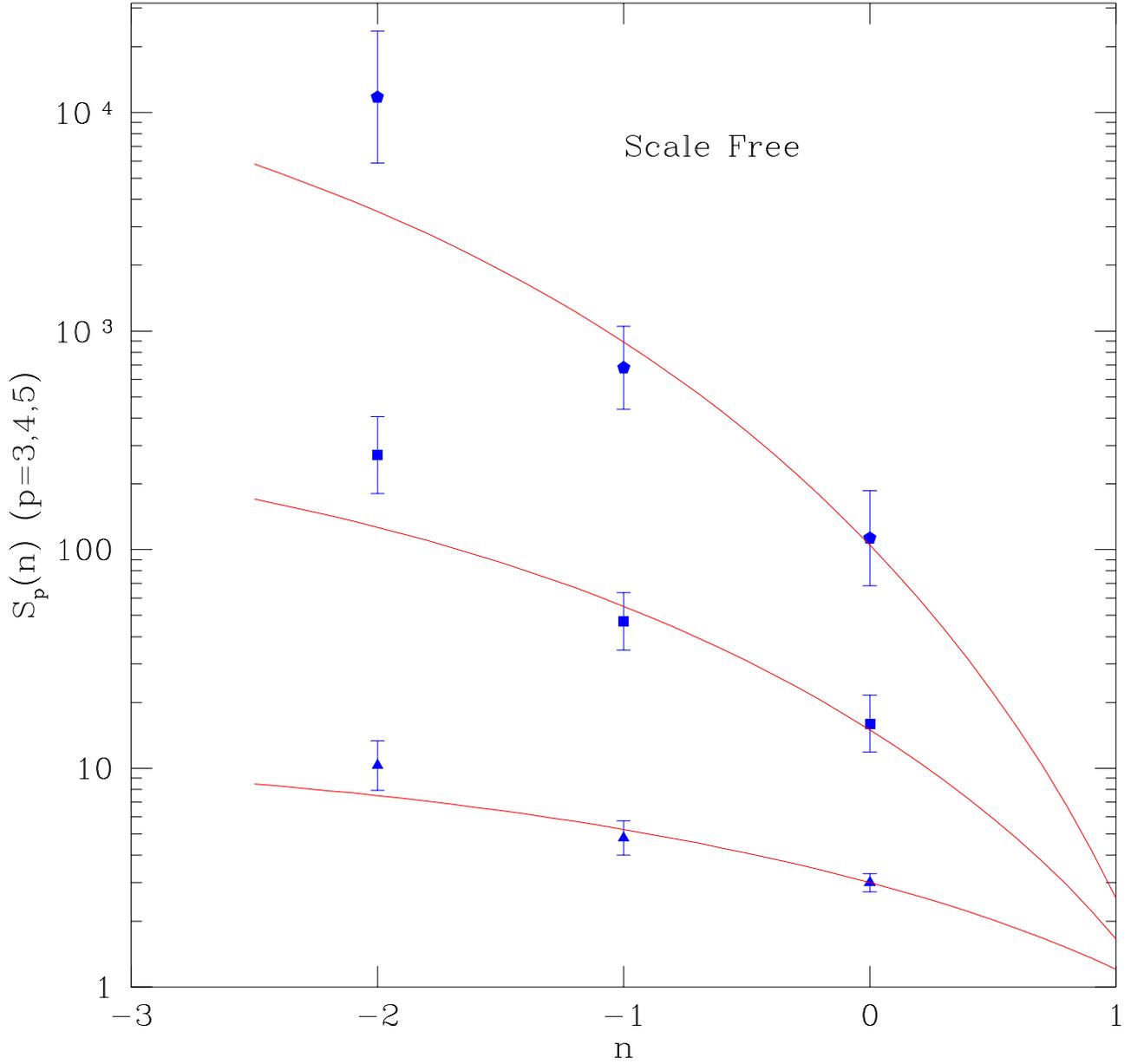}}

\caption{The hierarchical $S_p$ parameters: skewness (triangles,
$p=3$), kurtosis (squares, $p=4$), and pentosis (pentagons, $p=5$) are
shown for scale-free initial conditions in numerical simulations by
Colombi, Bouchet \& Hernquist (1996). The solid curves denote the
predictions of HEPT, Eqs.~(\protect\ref{s3}), (\protect\ref{s4}), and
(\protect\ref{s5}).}
\label{fig2}
\end{figure}

\begin{figure}[t!]
\centering
\centerline{\epsfxsize=18truecm\epsfysize=18truecm\epsfbox{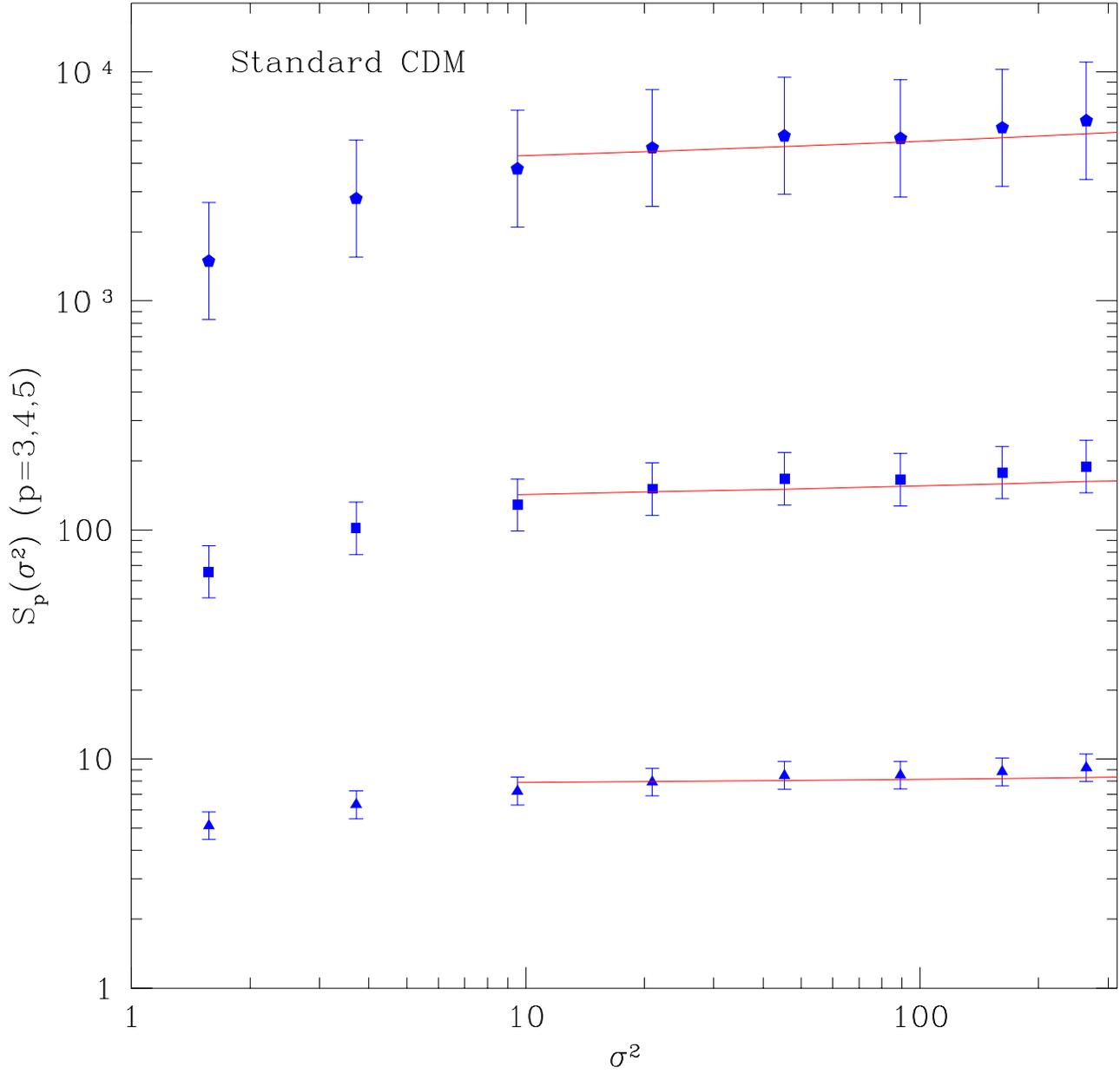}}

\caption{The skewness (triangles), kurtosis (squares), and pentosis
(pentagons) parameters for the standard CDM spectrum ($\sigma_8=0.34$)
from numerical simulations by Colombi, Bouchet, \& Schaeffer
(1994). The solid curves denote the predictions of HEPT,
Eqs.~(\protect\ref{s3}), (\protect\ref{s4}) and (\protect\ref{s5}).}
\label{fig3}
\end{figure}

\begin{figure}[t!]
\centering
\centerline{\epsfxsize=18truecm\epsfysize=18truecm\epsfbox{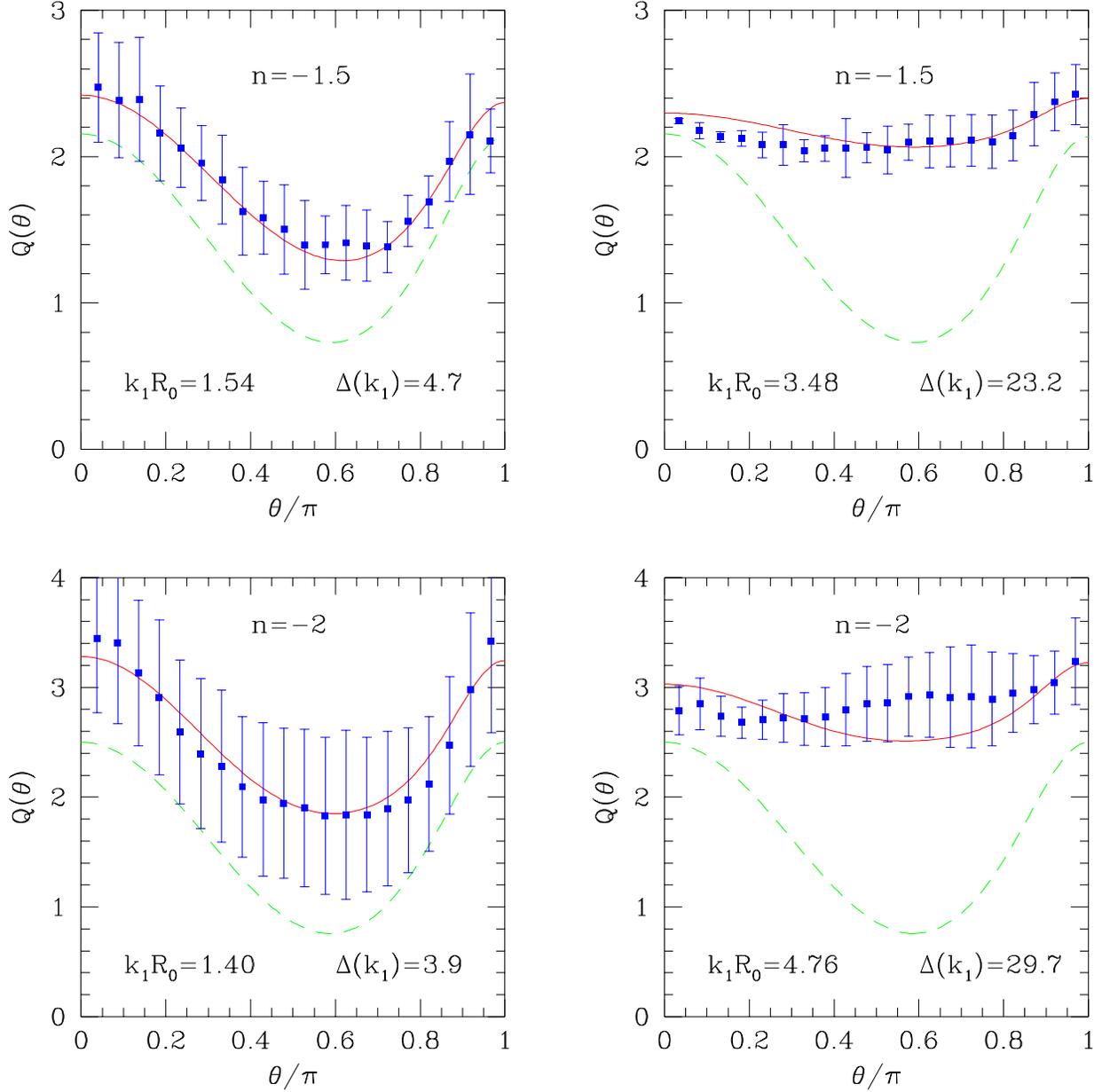}}

\caption{The three-point hierarchical Fourier amplitude $Q_3$ for
configurations with $k_1/k_2=2$ as a function of the angle $\theta$
between $\k_1$ and $\k_2$, at different scales for $n=-1.5$ (top) and
$n=-2$ (bottom) scale-free initial conditions. Symbols denote
measurements in numerical simulations (taken from \protect
\cite{SCFFHM98}, Figs.~1, 2, and 3). The solid curves show the
bispectrum fitting formula, Eqs.~(\protect\ref{Bfit}) to
(\protect\ref{coefit3}).  The dashed curves correspond to the
predictions of tree-level PT.}
\label{fig4}
\end{figure}

\begin{figure}[t!]
\centering
\centerline{\epsfxsize=18truecm\epsfysize=18truecm\epsfbox{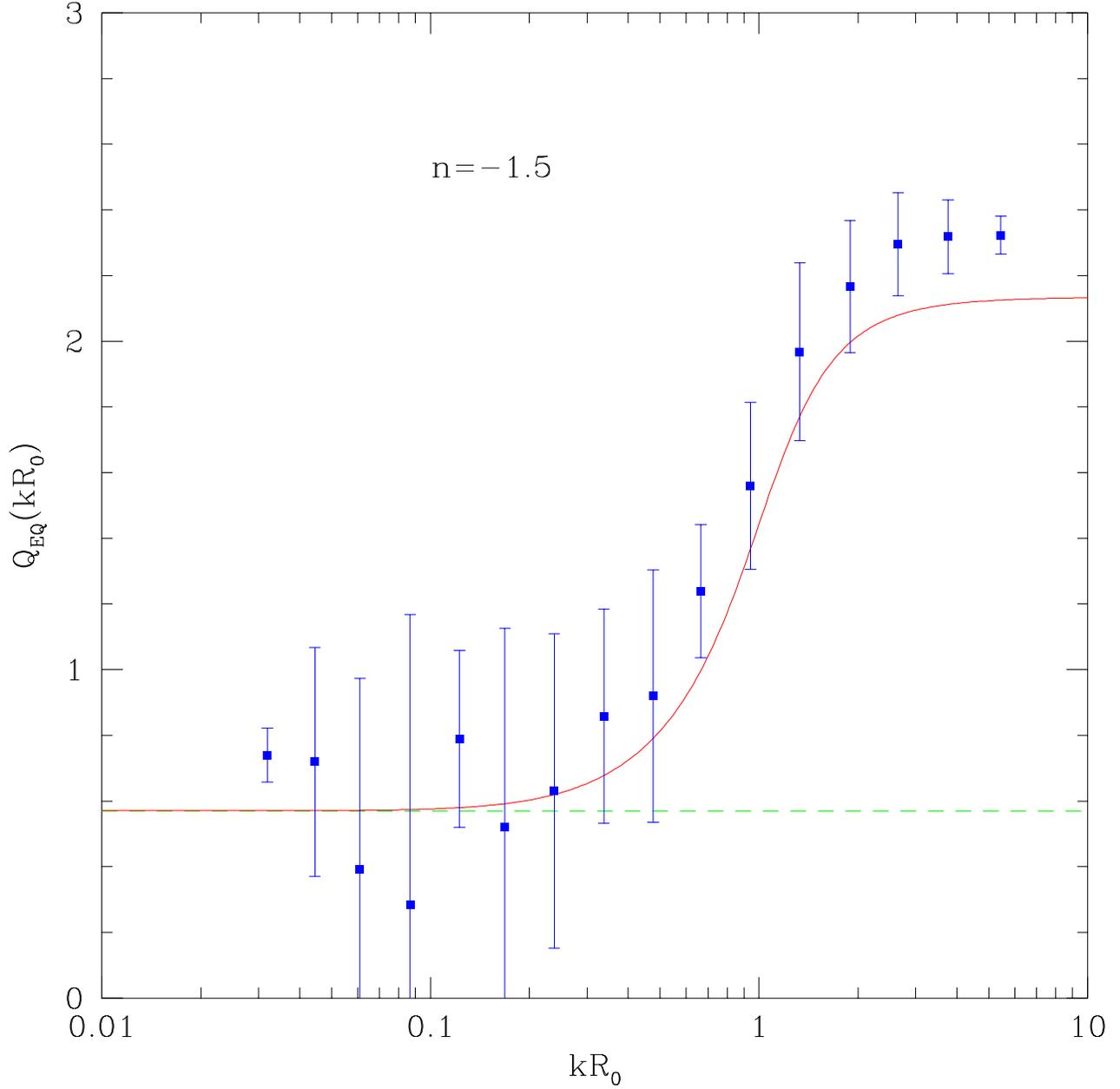}}

\caption{The three-point hierarchical Fourier amplitude $Q_{\rm EQ}$
for equilateral triangle configurations and $n=-1.5$ scale-free
initial conditions as a function of scale. Symbols denote measurements
in numerical simulations (taken from \protect \cite{SCFFHM98},
Fig.~9). The solid curve shows the bispectrum fitting formula,
Eqs.~(\protect\ref{Bfit}) to (\protect\ref{coefit3}), and the dashed
line corresponds to the prediction of tree-level PT, $Q_{\rm
EQ}=4/7$.}
\label{fig5}
\end{figure}

\begin{figure}[t!]
\centering
\centerline{\epsfxsize=18truecm\epsfysize=18truecm\epsfbox{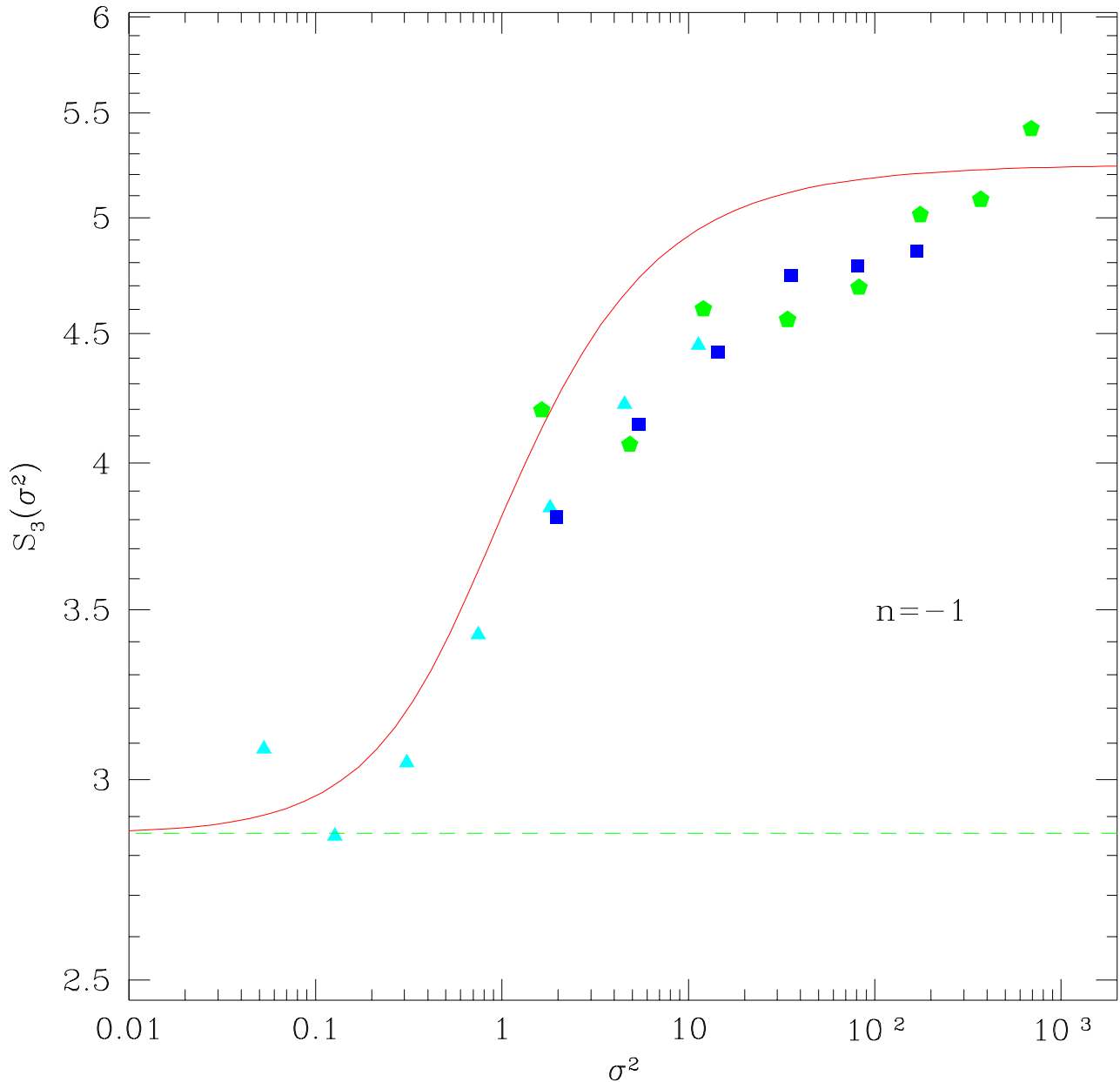}}

\caption{Predictions of our fitting formula for the skewness $S_3$ for
$n=-1$ scale-free initial conditions as a function of the variance
$\sigma^2$ for top-hat smoothing. Different symbols denote numerical
simulation measurements at different expansion factors, $a=2.5$
(triangles), $a=6.4$ (squares), and $a=16$ (pentagons), where initial
conditions were set at $a=1$. Error bars, not shown to avoid clutter,
are of the order of $20\%$ (\protect\cite{CBH96}). The solid curve
shows the fitting formula of Eq.~(\protect\ref{S3fit}), and the dashed
line corresponds to the prediction of tree-level PT, $S_3=20/7$. The
systematic underestimate of N-body results is likely due to transients
from initial conditions (see text).}
\label{fig6}
\end{figure}

\end{document}